# Determination of Carrier Polarity in Fowler-Nordheim Tunneling and Evidence of Fermi Level Pinning at the Hexagonal Boron Nitride / Metal Interface

*Yoshiaki Hattori*[*†], *Takashi Taniguchi*[‡], *Kenji Watanabe*[‡] *and Kosuke Nagashio*[**†§]

[†]Department of Materials Engineering, The University of Tokyo, Tokyo 113-8656, Japan
[‡]National Institute of Materials Science, Ibaraki 305-0044, Japan
[§]PRESTO, Japan Science and Technology Agency (JST), Tokyo 113-8656, Japan
[*]hattori@adam.t.u-tokyo.ac.jp, [**]nagashio@material.t.u-tokyo.ac.jp

**KEYWORDS:** *h*-BN, Fowler-Nordheim tunneling current, Fermi level pinning, pinning factor, barrier height.

**ABSTRACT:** Hexagonal boron nitride (*h*-BN) is an important insulating substrate for two-dimensional (2D) heterostructure devices and possesses high dielectric strength comparable to $SiO_2$. Here, we report two clear differences in their physical properties. The first one is the occurrence of Fermi level pinning at the metal/*h*-BN interface, unlike that at the metal/$SiO_2$ interface. The second one is that the carrier of Fowler-Nordheim (F-N) tunneling through *h*-BN is a hole, which is opposite to an electron in the case of $SiO_2$. These unique characteristics are verified by *I-V* measurements in the graphene/*h*-BN/metal heterostructure device with the aid of a numerical simulation, where the barrier height of graphene can be modulated by a back gate voltage owing to its low density of states. Furthermore, from a systematic investigation using a variety of metals, it is confirmed that the hole F-N tunneling current is a general characteristic because the Fermi levels of metals are pinned in the small energy range around ~3.5 eV from the top of the conduction band of *h*-BN, with a pinning factor of 0.30. The accurate energy band alignment at the *h*-BN/metal interface provides practical knowledge for 2D heterostructure devices.

## INTORODUCTION

Hexagonal boron nitride (*h*-BN) is an indispensable layered material used as an insulating film in two-dimensional (2D) heterostructure electronic devices because of its potential characteristics. It has a wide bandgap ($E_g$ = ~6 eV),[1] high thermal stability, an atomically flat surface, ideally no dangling bonds, and quite a few charge traps. Although several oxide insulating nanosheets have attracted attention recently, *h*-BN is the only layered insulator with covalent bonds closed in-plane. Therefore, its high dielectric strength and electrical reliability have been reported to be comparable to $SiO_2$.[2] In heterostructure devices, the understanding of the hetero-interface with *h*-BN, particularly the energy band alignment, is critical for designing the electronic device components. With regards to the energy band alignment at interfaces of 2D materials and metals, Fermi level pinning has recently been reported for semiconducting 2D materials such as $MoS_2$ or $MoTe_2$,[3,4] similarly to conventional semiconductors. However, little systematic study has been conducted for *h*-BN even though *h*-BN with a metallic gate electrode is often used in 2D heterostructure field-effect transistors (FETs), where the band alignment directly affects threshold voltage shift through flat band voltage.

Regarding the measurement of the barrier height for the band alignment, it can be estimated from the current-voltage (*I-V*) characteristic electrically. For the semiconductor/metal case, the dominant current is generally a thermionic emission current because of its relatively small barrier height. Therefore, the barrier height is estimated from the temperature dependence of the thermionic emission current.[4] On the other hand, the large barrier height at the insulator/metal interface prevents the thermionic emission current from being detected. Therefore, carrier injection into insulator techniques such as tunneling,[5] avalanche carrier,[6] internal photoemission,[7] and high-energy radiation[8] are necessary. Here, a common problem is whether the estimated barrier height is for electrons or for holes, because carriers with opposite signs are, in principle, simultaneously injected at the cathode for electrons and the anode for holes. The carrier polarity depends on the magnitude of each barrier height. In this way, the determination of the polarity of the transported carriers in an insulator is key to understanding accurate band alignment.

So far, the barrier height (2.4–4.2 eV[9, 10, 11]) at the $SiO_2$/metal interface has been well understood as being for electrons, while the barrier height at the $h$-BN/metal interface has been estimated to be 2–3 eV[12–14] and treated similar to that for electrons, without discussing the carrier polarity. Therefore, a careful investigation is required because the estimated value is close to the half of the bandgap of $h$-BN (6 eV).[1] In addition, from the viewpoint of reliability on dielectric strength or film quality evaluation in insulators, the determination of the polarity is quite important to identify the energy levels for many kinds of defects in the film, which cause the carrier trap and lead to dielectric breakdown.[11]

In this study, the use of a monolayer graphene electrode is proposed to clarify the polarity of a Fowler-Nordheim (F-N) tunneling current injected in $h$-BN, as shown in **Figure 1a**. The barrier height of graphene can be modulated by the back gate voltage

($V_{BG}$) due to the low density of states (*Dos*) in graphene. The dependence of the *I-V* characteristic on $V_{BG}$ is classified into six types in **Figure 1b** based on the relationships among the position of the Fermi levels of metal ($\Phi_{bm}$) and graphene ($\Phi_{bg}$) and the midgap of $h$-BN ($E_g/2$). Once both the carrier polarity and barrier height for a certain metal are revealed exactly by using the graphene electrode, all the information on the band alignment for a variety of metals can be simply determined by an F-N plot analysis in a conventional metal/$h$-BN/metal structure device. Therefore, the polarity of F-N tunneling current was determined first. Then, the band alignment for a variety of metals was systematically investigated. Finally, the Fermi level pinning was discussed.

## EXPERIMENTAL SECTION

In this study, the graphene/$h$-BN/metal devices were fabricated from exfoliated Kish graphite and $h$-BN which is grown by a high-pressure and high-temperature method[15]. The exfoliated monolayer graphene was transferred onto a $SiO_2$(~90 nm)/$n^+$-Si substrate, which is identified by an optical microscope (LV100, Nikon) equipped with a camera (DP71, Olympus), using subtle color differences. The thin $h$-BN flakes with a thickness of typically 10 nm were also prepared by a mechanical exfoliation technique and transferred to a poly(methyl methacrylate) (PMMA) layer with a thickness of 1 μm, which was spin-coated on a polydimethylsiloxane (PDMS) sheet. Then, $h$-BN on the PMMA/PDMS sheet was placed on the graphene using the alignment system.[16] After the removal of the sacrificial PMMA layer with acetone and isopropyl alcohol, the top electrodes were patterned on $h$-BN by electron beam (EB) lithography. Prior to metal vacuum deposition for the top electrodes (50-nm Au / 15-nm Cr), ozone treatment was performed for 5 min to remove the PMMA residue on the $h$-BN surface. Alternatively, the fabrication procedure of the metal/*h*-BN/metal devise is the same as that for the graphene device except for the use of a bottom electrode (30-nm Au / 5-nm Cr), which was fabricated on thermal oxidized silicon by EB lithography.

The electrical measurements were performed in a vacuum (~5.0×10⁻³ Pa) at room temperature (21–25 °C) using a semiconductor device parameter analyzer (B1500A, Keysight Technology) for *I-V* and a LCR meter (E4980A, Keysight Technology) for *C-V*. The magnified device image was taken by SEM (JSM-6010, JEOL) with a 5-kV accelerating voltage.

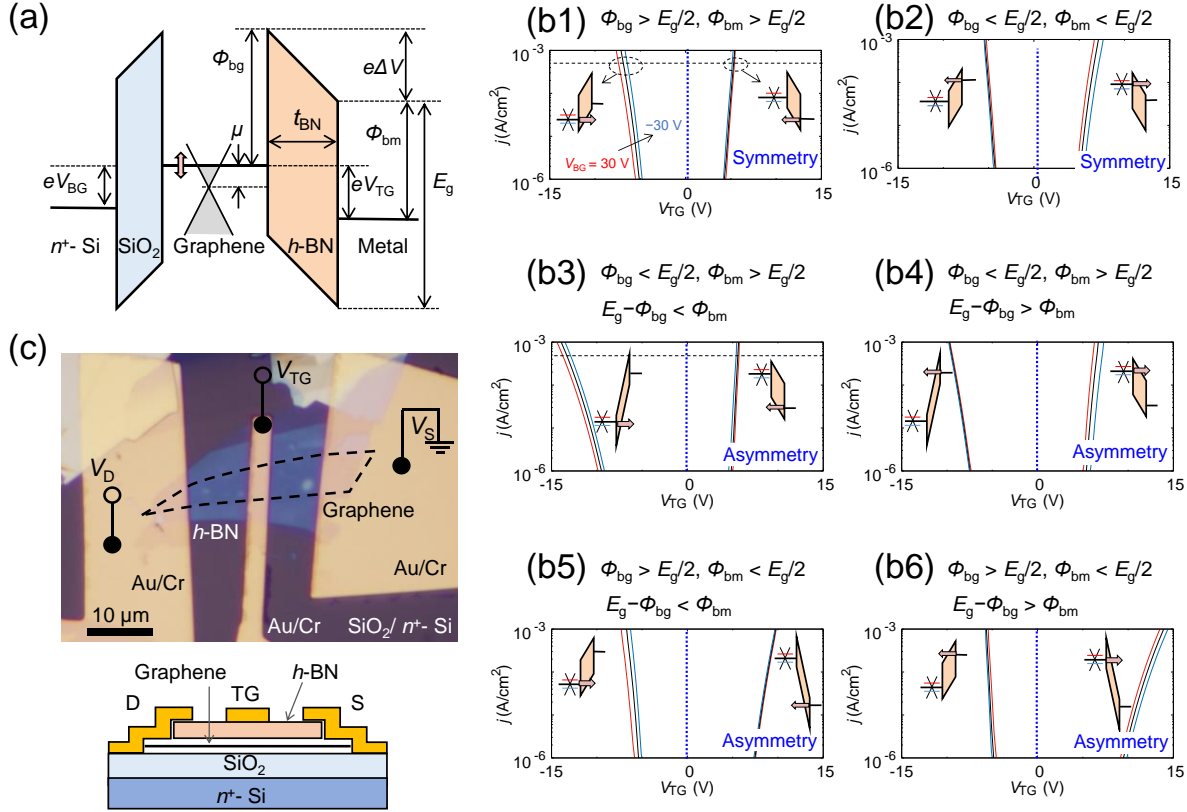

**Figure 1**: **(a)** Schematic band diagram for graphene/*h*-BN/metal structure. **(b1–6)** *I-V*$_{TG}$ characteristics of six different cases, which are classified by the relationships between the positions of the Fermi levels of metal and graphene and the midgap of *h*-BN. The two insets in each figure show the schematic band diagram at *V*$_{TG}$ applied to obtain the same current. For simplicity, we restrict the discussion to only the dominant polarity of the current here. In the case of b1, the carrier polarity of the F-N tunneling from both graphene and metal is positive, where the large change of F-N tunneling current appears in the negative voltage side since the barrier height for carrier injection from graphene is directly changed by modulating the Fermi level in graphene. When positive *V*$_{TG}$ is applied, the modulation of the Fermi level in graphene only alters the effective voltage stress in *h*-BN (Δ*V*). Therefore, the change in the F-N tunneling current is small. For b2, the situation is completely opposite to that of b1. In the case of b3, the carrier polarity from both graphene and metal is positive. However, a larger negative *V*$_{TG}$ is required to detect the same current because the barrier height for current injection from graphene is much larger than that in b1. As a result, the *I-V*$_{TG}$ becomes asymmetric. A similar asymmetry of *I-V*$_{TG}$ is seen in b4, where the carrier polarities from both graphene and metal are negative. The difference in b3 and b4 is the *V*$_{BG}$ dependence of the F-N tunneling current, because the change in barrier height from modulating the Fermi level of graphene is only effective when the F-N tunneling occurs from the graphene side. b5 and b6 are opposite situations to b4 and b3, respectively. It should be noted that the carrier polarity in F-N tunneling can be determined by these six classifications of *I-V* characteristics in any situation. **(c)** Optical microscope image and the cross sectional schematic drawing of the typical graphene/*h*-BN/metal device.

## RESULTS AND DISCUSSION

**Figure 1c** presents the optical microscope image and the cross-sectional schematic drawing for the typical graphene/*h*-BN/metal device. This device can be regarded as two different structures; one is the metal/insulator/metal (MIM) structure and the other is the graphene FET structure with dual-gates of *h*-BN and SiO$_2$ insulators. In the MIM structure, graphene is treated as the electrode and the Fermi-level of graphene can be modulated by the back gate. The overlap area for the graphene and top metal electrode is typically 25 μm$^2$. In the electrical measurement on MIM, graphene in the region without top metal electrode can be treated as leads. On the other hand, the graphene FET can be used to quantitatively estimate the Fermi level as well as the Dirac point of graphene.

**Figure 2a** shows the *I-V* characteristic of the MIM structure, that is, the graphene/*h*-BN/Cr/Au device for different $V_{BG}$. The top gate voltage ($V_{TG}$) was applied with respect to the sourse electrode gradually from 0 V to the negative side with a ramp speed of 0.2 V/s while keeping $V_{BG}$ constant. In the measurements, the drain electrode was floated. The applied $V_{TG}$ is terminated manually when the current density reaches around ~0.6 × 10$^{-5}$ A/cm$^2$. The series of measurements was repeated with different $V_{BG}$ from 30 V to −30 V with a step of 5 V to modulate the Fermi level of graphene. It should be emphasized that leakage current through the back gate SiO$_2$ insulator was not detected. Then, the positive side of $V_{TG}$ was also measured in the same manner. F-N tunneling currents were detected for $|V_{TG}| > 6$ V for all the cases, since linear relations were clearly found in the F-N plots (**Figure S1a**). As shown in **Figure 2a**, it is clear that the $V_{TG}$ dependence on F-N tunneling currents only appeared on the negative side, not on the positive side. The comparison of this behavior with six kinds of *I-V* characteristics in **Figure 2a** suggests that the present situation is that of case (b1), that is, both F-N tunneling currents observed for positive and negative $V_{TG}$ are indeed hole currents. In other words, both Fermi levels are below the midgap of *h*-BN. The modulation of barrier height between graphene and *h*-BN by $V_{BG}$ changes the tunneling distance for holes, which explains the observation of $V_{TG}$ dependence on F-N tunneling currents only on the negative $V_{TG}$ side. This is not the case for the positive $V_{TG}$ side.

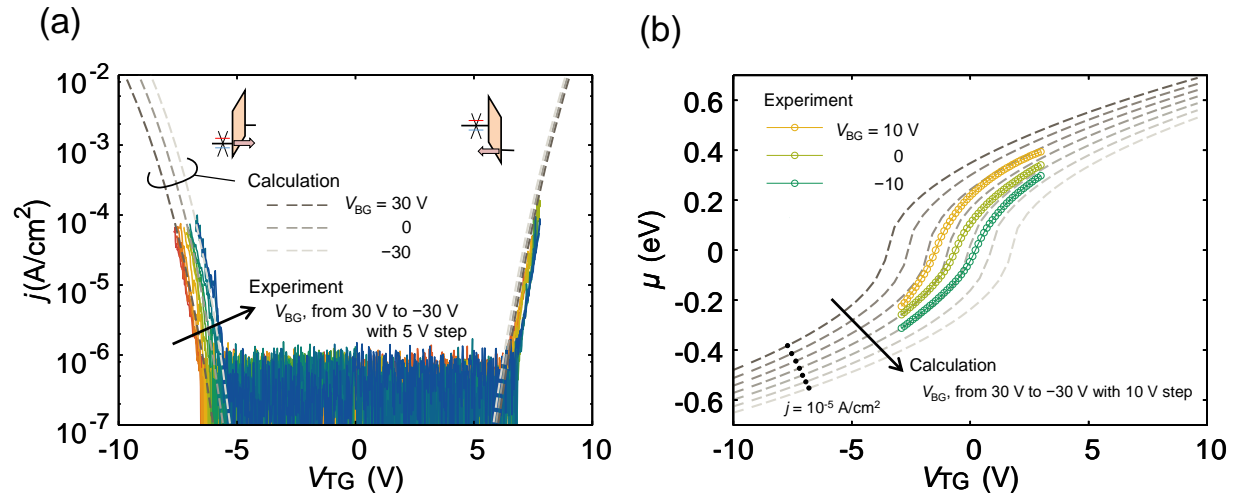

**Figure 2**: **(a)** F-N tunneling current as a function of $V_{TG}$ for different $V_{BG}$. The numerical calculation based on Equation (5) corresponds well with the experimental data, supporting the hypothesis of the hole current. **(b)** Fermi levels in the graphene modulated by $V_{TG}$. The numerical simulation is in good agreement with the experimental values extracted from *C-V* measurement.

Next, we should verify how the Fermi level in graphene is experimentally modulated by $V_{TG}$ and $V_{BG}$. When graphene is used as one side of the electrodes for *h*-BN and $V_{TG}$ is applied, the effective voltage applied to *h*-BN is intrinsically reduced from $V_{TG}$ due to the extra kinetic energy required to induce carriers in the graphene with the low density of states ($Dos_g$).[17] Therefore, the total capacitance ($C_{total}$) can be described by $1/C_{total} = 1/C_{BN} + 1/C_Q$, where $C_{BN}$ and $C_Q$ are the geometric capacitance of *h*-BN and the quantum capacitance ($C_Q = e^2 Dos_g$), respectively. Moreover, the Fermi energy in graphene is additionally modulated by $V_{BG}$ in the above experiment. Here, we show the brief summary of the experiments in which the device in **Figure 1c** is dealt as the graphene FET with *h*-BN top gate insulator. Both *I*-$V_{TG}$ and capacitor-voltage (*C*-$V_{TG}$) measurements were conducted in the range of −3 V to 3 V for $V_{TG}$ for different $V_{BG}$, as shown in **Figure S2a-b**. Dirac point voltages ($V_{DP}$) for a top-gate sweep were plotted as a function of $V_{BG}$, as shown in **Figure S2c**. The slope provides the coupling ratio of the geometric capacitances of the *h*-BN and SiO₂ layers. Then, $C_{BN}$ can be calculated to be 0.45 µF/cm², if we use the thickness of *h*-BN as measured by the atomic force microscope (AFM) (9 nm) and the thickness of SiO₂ measured by the ellipsometer (89 nm). $C_Q$ in graphene can be calculated from the *C*-$V_{TG}$ measurement using $C_{BN}$ after the parasitic capacitance ($C_{para}$) is removed properly. Moreover, the relation of $V_{TG}$' and modulated energy of Fermi level in graphene ($\mu$) is calculated using the following equation:

$$\mu = e\left(V_{TG}' - \int_0^{V_{TG}'} \frac{C_{total}}{C_{BN}} dV_{TG}'\right),$$ (1)

where $V_{TG}$' is defined as $V_{TG}' = V_{TG} - V_{DP}$. The detailed analysis can be found in our previous paper.[17] **Figure S2f** shows $C_Q$ extracted experimentally as a function of $\mu$ for $V_{BG} = -10$ V, suggesting that $C_Q$ has been accurately calculated. The $\mu$ in graphene is again plotted as a function of $V_{TG}$ for different $V_{BG}$ in **Figure 2b**. When $V_{TG}$ was applied to −3 V at $V_{BG} = 0$ V, $\mu$ is modulated to 0.27 eV and further extended to 0.32 eV by applying $V_{BG} = -10$ V. It is clear that $V_{BG}$ can additionally control $\mu$ in graphene, which results in the change in $V_{TG}$ required for the detection of the F-N tunneling current, as shown in **Figure 2a**.

Here, the $\mu$ modulation observed in the graphene FET with the *h*-BN top gate insulator is compared with the simulation results. $\mu$ can be related to the carrier density of graphene ($n_g$) by the following series of equations:[18–20]

$$\varepsilon_{ox} F_{ox} + \varepsilon_{BN} F_{BN} = n_g e,$$
$$V_{TG} = e\Delta V + \mu + \Phi_{bm} - \Phi_{bg},$$
$$\mu = \hbar v_F \sqrt{\pi n_g},$$

where $\Phi_b$ is the barrier height, $\hbar$ is Planck's constant, $v_F$ is the Fermi velocity of electrons in graphene, and $\varepsilon$ is the dielectric constant. The subscripts ox, BN, g, and m are attached to the properties of SiO₂, *h*-BN, graphene, and metal, respectively. $\Delta V$ is defined as the effective voltage stress in *h*-BN and is shown graphically in **Figure 1a**. $F_{ox}$ and $F_{BN}$ are the effective electric fields and are defined as $F_{BN} = \Delta V/t_{BN}$ and $F_{ox} = V_{BG}/t_{ox}$, respectively, where *t* is the thickness of each insulating film. In the simulation, the residual carrier density[17] is assumed to be zero. The experimentally obtained physical values for $\varepsilon_{BN}$, $t_{BN}$ and $t_{ox}$ were used in this simulation. These equations were numerically solved by the bisection method. The results for $\mu$ calculated as a function of $V_{TG}$ for different $V_{BG}$ are compared with the experimental data in **Figure 2b**, indicating good agreement with the experiment. In addition, the numerically calculated $V_{DP}$ obtained from $d\mu/dV_{TG}$ also matches the experimental results (**Figure S2c, g**).

Next, the change in F-N tunneling current through $h$-BN due to the modulation of $\mu$ in graphene by $V_{BG}$ is further proven by the following numerical calculation. By using the calculated $\mu$, the tunneling current was estimated using Bardeen's tunneling equation.[21]

$$j = A \int Dos_m(E)\, Dos_g(E)\, T(E)[f(E - eV_{TG}) - f(E)]dE,$$

where the constant $A$ is treated as a fitting parameter, and $j$, $E$, $T$, and $Dos_m$ are the current density, energy, transmission probability, and density of states of the metal, respectively. The function $f$ is a Fermi-Dirac distribution. $Dos_g$ is calculated by the following equation:[22, 23]

$$Dos_g(E) = \frac{2|E - E_{DP}|}{\pi(h\nu_F)^2},$$

where $E_{DP}$ is the energy difference based on the Dirac point. For simplification, constant $Dos_m$ was assumed,[21] and the dispersion relation of graphene was treated as an isotropic parabolic energy band.[24] Then, $T(E)$ is expressed as follows by applying the WKB approximation:[25]

$$T(E) = exp\left(-2\int_0^{t_{BN}} \left(\frac{2m^*}{h^2}eV(x)\right)^{\frac{1}{2}} dx\right),$$

where $x$ is the distance from the interface at the anode and $eV(x)$ is the shape of the potential barrier for holes, which can be calculated by Equations (2-4). $m^*$ is the effective mass of an electron in $h$-BN and set to be 0.5 $m_0$, where $m_0$ is the mass of an electron in vacuum.[26] The calculated values of $j$ are shown in **Figure 2a** with dotted lines. The numerical calculations reproduced the experimental results well. In addition, it should be noted that the I-V characteristics for six cases in **Figure 1b** were also obtained from the numerical calculation for typical barrier heights. These results support the hypothesis of hole injection from the metallic electrode. Moreover, the $\mu$ in graphene, which leads to the F-N tunneling current of $10^{-5}$ A/cm$^2$ for the negative $V_{TG}$ side, is plotted as black circles in **Figure 2b**. It is clear that the $\mu$ in graphene is modulated additionally by ~0.2 eV for the $V_{BG}$ range of −30 to 30 V. This simulation is quite useful because the $\mu$ in graphene cannot be experimentally extracted during F-N tunnel current measurement through the graphene/$h$-BN/Cr/Au structure.

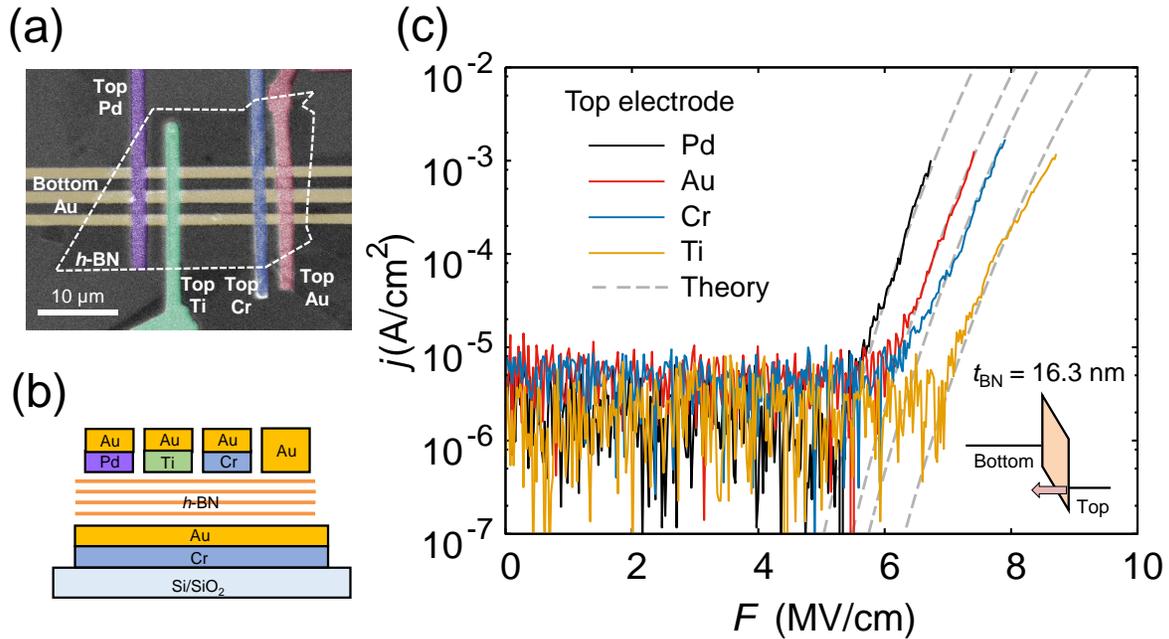

**Figure 3**: **(a)** Color-enhanced SEM image for the typical device. **(b)** Schematic drawing of the cross-sectional device structure. **(c)** F-N tunneling current as a function of electric field for different top electrodes. The electric field ($F$) is simply calculated by dividing the applied voltage by $t_{BN}$. The results of the numerical simulations are also plotted as broken lines.

Next, the barrier heights of a variety of metals for $h$-BN were investigated in metal/$h$-BN/metal structure devices. The color-enhanced scanning electron microscope (SEM) for the typical device is shown in **Figure 3a**, where four kinds of top electrodes are fabricated on the same single-crystal $h$-BN flake with an array structure (**Figure 3b**). The device was fabricated in a similar way to the graphene device, except for the use of Au(30 nm)/Cr(5 nm) bottom electrodes using $h$-BN with a thickness of ~16 nm. The device structure enables us to systematically evaluate the differences between the top electrode metals in F-N tunneling under the same experimental conditions. First, the $I$-$V$ test was conducted in a Cr/$h$-BN/Au device, in which the bottom Au electrode was always grounded.

The $I$-$V$ characteristics observed for positive and negative voltage sweeps show a symmetrical shape, as shown in **Figure S3**, suggesting that the Cr/$h$-BN/Au device is of the type (b1) or (b2) in **Figure 1**. This means that it has carriers of the same sign in its F-N tunneling current. Because the position of the Fermi level of Cr has been found to be below the midgap of $h$-BN, it can be concluded that the Fermi level of the bottom Au electrode is also below the midgap, that is, the type (b1) is the case. **Figure 3c** shows the $I$-$V$ characteristics for the positive voltage in a variety of metals. The measured current fits well with the theoretical F-N tunneling current, and all the barrier heights extracted by the F-N plots are smaller than the midgap energy of 3 eV when the band gap of $h$-BN is 6 eV.[1] Because the Au bottom electrode is common for all the top metal electrodes, by the same logic, the position of the Fermi level for the all tested metals is found to be below the midgap, i.e., the extracted barrier heights ($\Phi_b$') are for the hole injections. The $\Phi_{bm}$ (= $E_g$ − $\Phi_b$') of each metal, defined as the difference in energy between the conduction band minimum of $h$-BN and Fermi level of the metal, is summarized with respect to the work function of the metal ($\Phi_M$) in **Figure 4a**. A metal with a higher work function has a higher barrier height, which is a reasonable trend. Furthermore, the pinning of the Fermi levels of the metals is clearly observed.

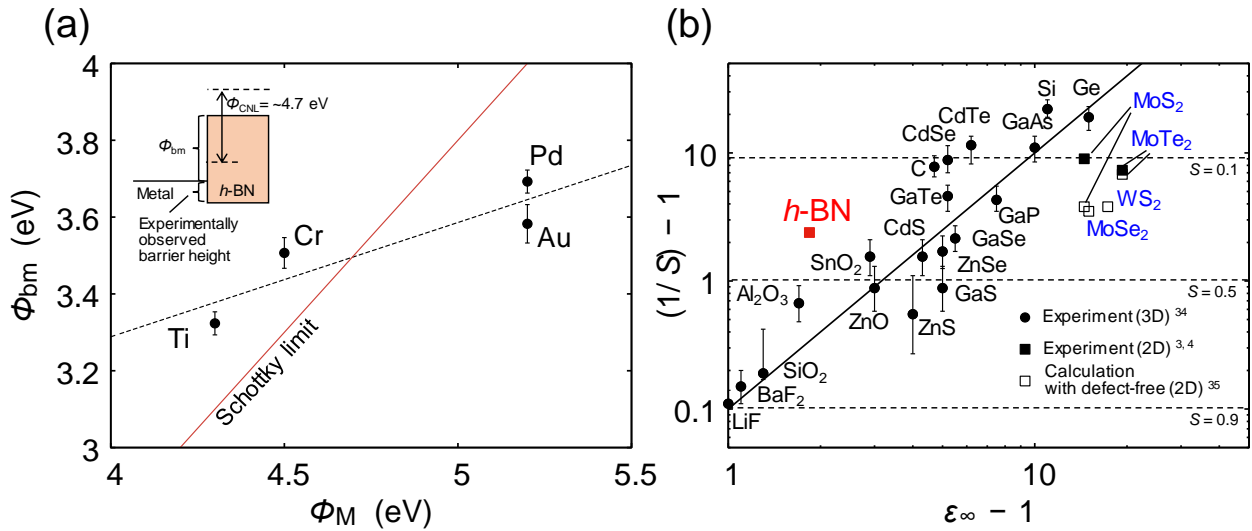

**Figure 4**: **(a)** $\Phi_{bm}$ as a function of the work function of metal, showing that a metal with higher work function has higher barrier height. The error bars in the figure are the standard deviation which were taken from more than 3 devices. **(b)** Comparison between various materials of 1/S−1 vs $\varepsilon_\infty$−1. Solid line indicates the empirical formula of $S = [ 1 + 0.1(\varepsilon_\infty - 1)^2) ]^{-1}$.

Through this study, the pinning effect was confirmed in the thickness range of 10–20 nm for $h$-BN, where the thickness dependence on the barrier height was not observed. On the other hand, for other 2D semiconductors, e.g., $MoS_2$, the barrier height for Au/$MoS_2$ interface depends on the layer number of $MoS_2$ and pinning effect, because the band gap of $MoS_2$ gradually changes with the layer number.[27] Similar situation is expected for $h$-BN since the band gap of $h$-BN is reported to decrease with increasing the layer number from monolayer to ~5 layers.[28] However, the present study is limited to the thick $h$-BN flakes.

Generally, the Fermi level pinning is characterized by $\Phi_{bm} = S\ (\Phi_M - \Phi_{CNL}) + (\Phi_{CNL} - \chi)$, where $\chi$ is electron affinity. $S$ ($= d\Phi_{bm}/d\Phi_M$) is the pinning factor, which is calculated from the slope in **Figure 4a**. This figure characterizes the degree of the pinning in a range of 0 to 1, where $S = 0$ and 1 represent an unpinned interface (Schottky limit) and full pinning (Bardeen limit), respectively. $\Phi_{CNL}$ is the charge neutrality level, which is defined as the energy from the vacuum level, and is determined from the intersection points with the Schottky limit ($S = 1$), as shown in **Figure 4a**. Therefore, $S$ and $\Phi_{CNL}$ for $h$-BN are calculated to be 0.30 and 4.7 eV, respectively. Note that $\chi = 1.2$ eV for the bulk case was used in this estimation.[29] Although the $\chi$ of $h$-BN under specific surface conditions has been known to be negative,[30] this is not the case for the present metal/$h$-BN/metal structure.

The origin of the Fermi level pinning has been debated for a long time. There are two well-known models; one uses intrinsic metal-induced gap states (MIGS).[31] The other uses extrinsic disorder induced gap states (DIGS),[32] where $S$ decreases with the increase of the defect states. Here, $S$ for $h$-BN is considered using the MIGS model, because the defect density in $h$-BN is relatively small ($10^9 - 10^{10}$ cm$^{-2}$).[33] In this model, $S$ is proposed to follow the empirical formula $S = [\ 1 + 0.1(\varepsilon_\infty - 1)^2)\ ]^{-1}$, where $\varepsilon_\infty$ is the optical relative dielectric constant.[34] **Figure 4b** shows the experimental values of $1/S - 1$ versus $\varepsilon_\infty - 1$ for various materials with the present $h$-BN data.[3, 4, 34] 2D materials, including $h$-BN, follow the general trend, although an $S$ value that is close to unity is intuitively anticipated for 2D materials because they are of a van der Waals nature without dangling bonds.[35, 36] In addition, the effect of extrinsic defects such as sulfur vacancies (~$10^{13}$ cm$^{-2}$) has been reported in $MoS_2$,[37] reflecting the difference between the experimental values and theoretical simulations in **Figure 4b**.[37, 38] Furthermore, it is interesting to consider the relation between the $S$ values and the crystallographic anisotropy of $h$-BN. Since $\varepsilon_\infty{}^{//c}$ is smaller than $\varepsilon_\infty{}^{\perp c}$ for $h$-BN,[39] $S_{\perp c}$ can be easily expected to be smaller than $S_{//c}$ owing to the strong covalent bonding in the end-bonded junction.[40] The present picture of the crystallographic anisotropy in $S$ is similar to that of previous research on semiconductor carbon nanotubes and 2D heterojunctions for the end-bonded junction and the planar junction.[41,42]

In addition, **Figure 4b** illustrates the unpinned interface for various oxide insulators. The thin oxide layer has been inserted between the semiconductor and the metal as a tunneling insulator to control the barrier height.[43, 44] It has been reported that the $MoS_2$/metal interface with strong pinning of $S$ = ~0.1 is weakened by inserting the thin $h$-BN.[45, 46] However, the pinning of $h$-BN itself was not discussed. Therefore, the present finding of the pinning in $h$-BN suggests a limited ability to control the band alignment, that is, the complete depining is not expected unlike oxide insulators.

## CONCLUSION

The carrier polarity of F-N tunneling through $h$-BN is verified as that of holes by (1) the modulation of $I$-$V$ characteristics by the back gate voltage in graphene/$h$-BN/metal heterostructure devices, (2) numerical simulations, and (3) the proportional relation between the barrier height and the work function of metal. In the course of a systematic study of various metals, the Fermi level pinning of metal with $h$-BN is clearly revealed. Because it is pinned in a small energy range of approximately 3.5 eV from the top of the conduction band of $h$-BN with a pinning factor of 0.30, the hole F-N tunneling current is a practically general characteristic for various metals. For dielectric breakdown, the $h$-BN is broken by the hole F-N tunneling current, unlike $SiO_2$, because the excessive F-N tunneling current leads to dielectric breakdown. The present findings of hole injection and the Fermi level pinning for $h$-BN provide a practical guideline in band alignment in 2D van der Waals heterostructure devices and help to clarify the mechanism of dielectric breakdown in $h$-BN.

## SUPPORTING INFORMATION

The F-N plots for a graphene/$h$-BN/metal device, analysis of $I$-$V$ and $C$-$V$ measurement for graphene/$h$-BN/metal device, $I$-$V$ characteristic for both polarity in Cr/$h$-BN/Au device. The Supporting Information is available free of charge via the Internet at http://pubs.acs.org.

## AUTHOR INFORMATION

### Corresponding Author
Email:        *hattori@adam.t.u-tokyo.ac.jp, **nagashio@material.t.u-tokyo.ac.jp

### Acknowledgement
This research was supported by the JSPS Core-to-Core Program, A. Advanced Research Networks, JSPS KAKENHI Grant Numbers JP17K14656, JP25107004, JP16H04343, JP16K14446, and JP26886003, and JST PRESTO Grant Number JPMJPR1425, and Mikiya Science and Technology Foundation, and Mizuho Foundation for the Promotion of Sciences.

### Notes
The authors declare no competing financial interests.

Contact by Using Atomic Thick h-BN as a Tunneling Layer, *Advanced Materials*, 28(37), 8302-8308.

**TOC graphic figure**

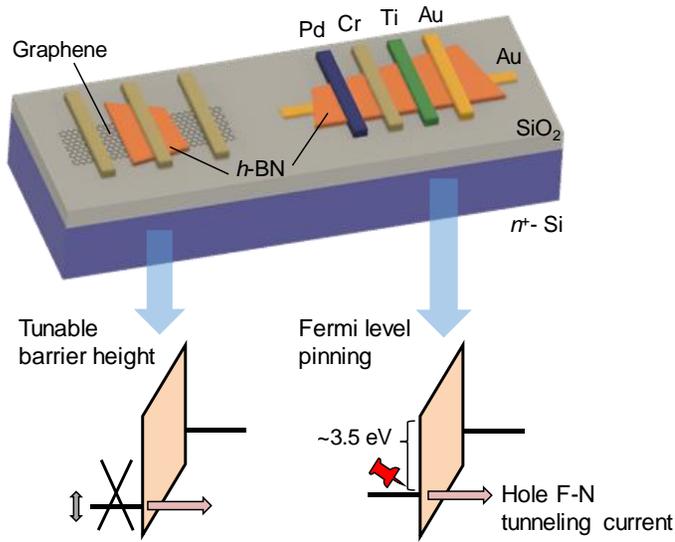

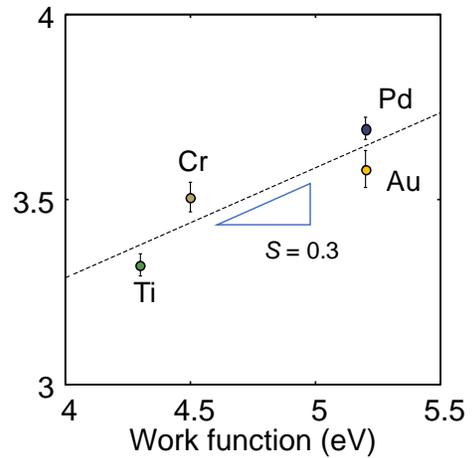

# Determination of Carrier Polarity in Fowler-Nordheim Tunneling and Evidence of Fermi Level Pinning at the Hexagonal Boron Nitride / Metal Interface

*Yoshiaki Hattori[\*†], Takashi Taniguchi[‡], Kenji Watanabe[‡] and Kosuke Nagashio[\*\*†§]*

[†]Department of Materials Engineering, The University of Tokyo, Tokyo 113-8656, Japan

[‡]National Institute of Materials Science, Ibaraki 305-0044, Japan

[§]PRESTO, Japan Science and Technology Agency (JST), Tokyo 113-8656, Japan

[\*]hattori@adam.t.u-tokyo.ac.jp, [\*\*]nagashio@material.t.u-tokyo.ac.jp

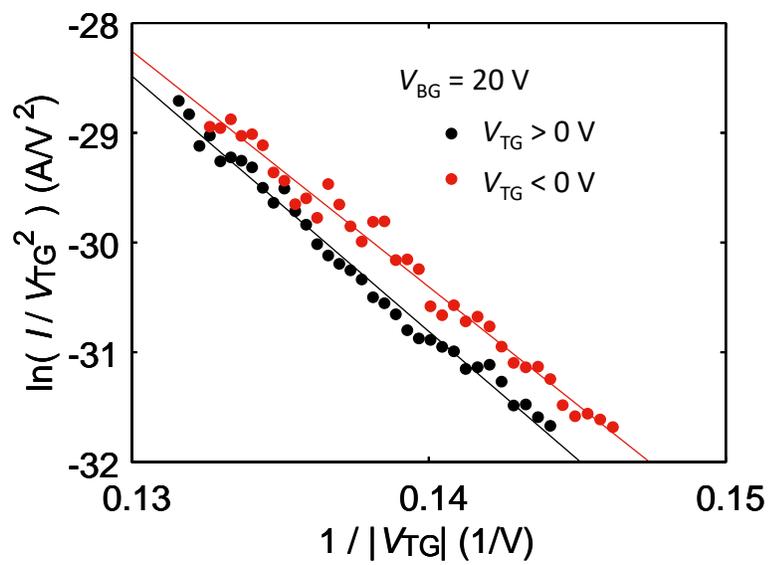

**Figure S1:** F-N plot for the graphene/*h*-BN/metal device. The linear relationship indicates that the dominant current is a F-N tunneling current.

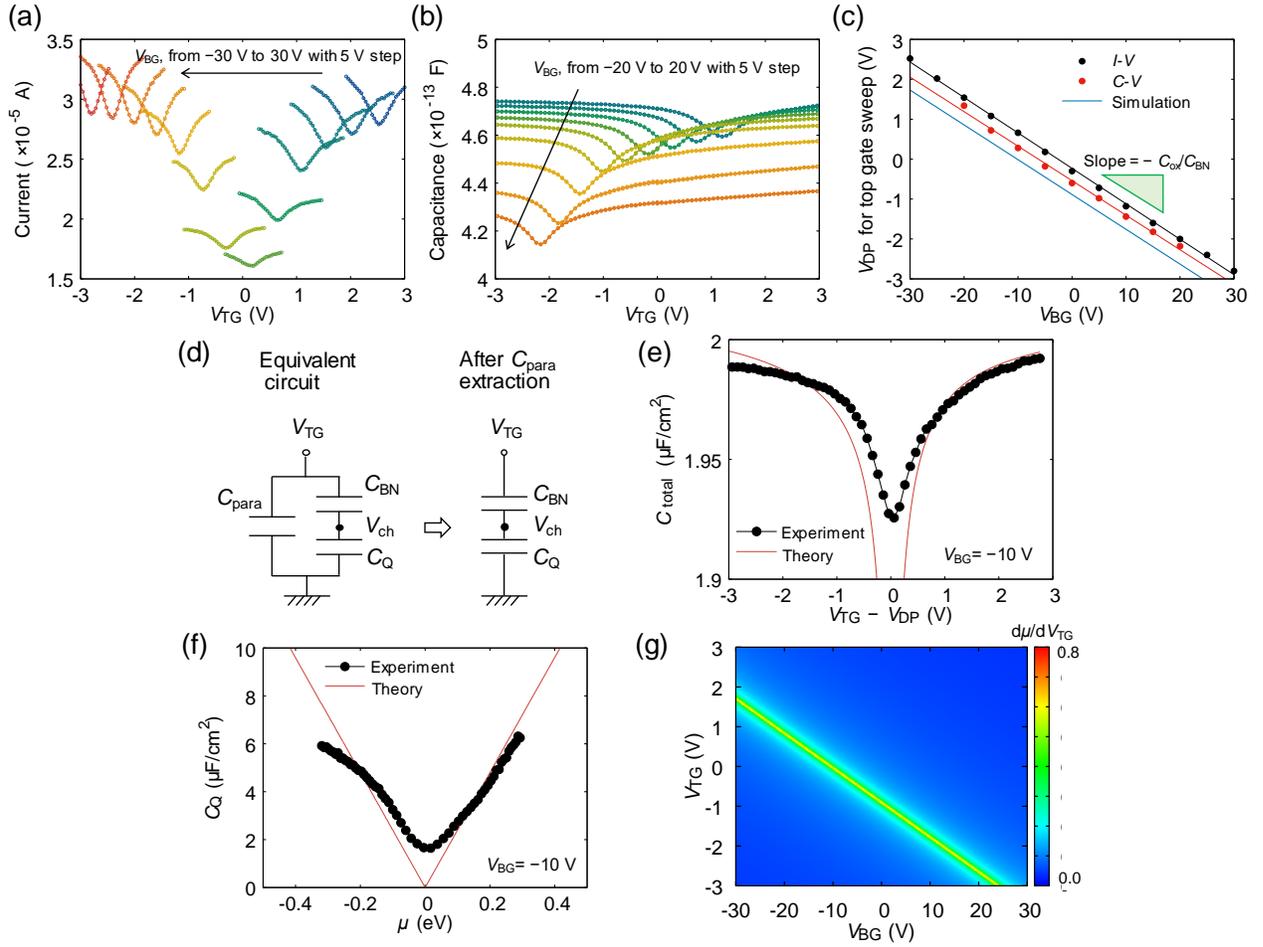

**Figure S2: (a)** $I$-$V_{TG}$ characteristics for different $V_{BG}$, where the region near the Dirac point is plotted. **(b)** $C$-$V_{TG}$ characteristics for different $V_{BG}$. **(c)** $V_{DP}$ for top gate sweep as a function of $V_{BG}$. The slope indicates the ratio of the geometrical capacitances of the $h$-BN and SiO$_2$ layers. The solid lines for $I$-$V$ and $C$-$V$ are fitted lines. The solid line in blue is a calculated value which is extracted from **(g)**. The $V_{DP}$ corresponds to the maximum value of d$\mu$/d$V_{TG}$ for constant $V_{BG}$. All slopes match well. **(d)** The equivalent circuit for the graphene FET with $h$-BN top gate. Then, after the extraction of parasitic capacitance ($C_{para}$), equivalent circuit can be simplified. **(e)** $C_{total}$ as a function of $V_{TG} - V_{DP}$ at $V_{BG} = -10$ V, where the parasitic capacitance ($C_{para}$) is removed properly by fitting with the equivalent circuit of **(d)**. **(f)** $C_Q$ as a function of $\mu$ at $V_{BG} = -10$ V. **(g)** Simulated color plot of d$\mu$/d$V_{TG}$ as a function of $V_{TG}$ and $V_{BG}$. The highest d$\mu$/d$V_{TG}$ at constant $V_{BG}$ corresponds to $V_{DP}$ since the *Dos* of graphene is minimal.

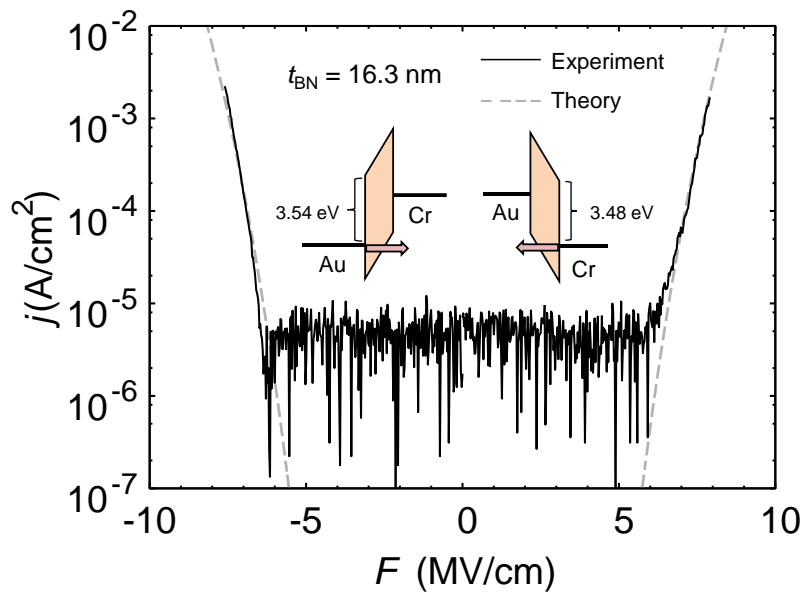

**Figure S3:** F-N current as a function of electric field for both polarities in a Cr/*h*-BN/Au device. The barrier height is estimated properly not only from the slope in the F-N plots but also by fitting to the analytical solution.